**Different Ways of Thinking about Street Networks and Spatial Analysis**

Bin Jiang[1], and Atsuyuki Okabe[2]

[1]Department of Technology and Built Environment, University of Gävle, Sweden
[2]School of Cultural and Creative Studies, Aoyama Gakuin University, Japan

Street networks, as one of the oldest infrastructures of transport in the world, play a significant role in modernization, sustainable development, and human daily activities in both ancient and modern times. Although street networks have been well studied in a variety of engineering and scientific disciplines, including for instance transport, geography, urban planning, economics, and even physics, our understanding of street networks in terms of their structure and dynamics remains limited, especially when dealing with such real-world problems as traffic jams, pollution, and human evacuations for disaster management. Thanks to the rapid development of geographic information science and its related technologies, abundant street network data have been collected to better understanding the networks' behavior, and human activities constrained by the networks. For example, OpenStreetMap has assembled hundreds of thousands of gigabytes of data for streets, and for other related geographic objects, such as public transports, building footprints, and points of interest. Given this context, we predict that, in the near future, increasing amounts of research will be published regarding the underlying structure and dynamics of street networks. This special issue has collected five of the best papers from 19 works submitted to and presented at the ICA workshop on street networks and transport (https://sites.google.com/site/icaworkshop2013/). Due to time constraints, we were unable to include a number of other high-quality papers, but we are confident that these papers will be added elsewhere in the literature soon.

One goal of this special issue is to promote different ways of thinking about and understanding street networks, and of conducting spatial analysis. Network spatial analysis involves a set of statistical and computational methods developed by Okabe and Sugihara (2012), as demonstrated again in Shiode and Shiode 2014, for analyzing events occurring along networks. The network spatial analysis clearly differs from conventional spatial analysis, which assumes a continuous Euclidean space rather than space constrained to networks. Current network analysis in geographic information systems (GIS) is essentially geometry oriented, so it is hard to address some research issues related to the underlying structure. In this regard, the topological representation to be introduced in the following text has enabled us to uncover the underlying scaling pattern of street networks. Four of the special issue papers (Gil 2014, Lerman et al. 2014, Mohajeri and Gudmundsson 2014, Wei and Yao 2014) have adopted or are well connected to the topological representation, which is powerful for understanding street hierarchies or geographic forms and processes in general.

Conventionally, geographic space is considered as a continuous Euclidian space that is divided or subdivided into different areas, which authorities often define and delineate administratively and legally. Data collected on geographic space are assigned into individual areas, and are therefore assumed to be homogenous in each area. There were good reasons for this assumption during the small-data era, when data was not as rich as it is in the big-data era. For example, for the purpose of protecting privacy, data are deliberately aggregated rather than analyzed at individual level; data are often estimated statistically or roughly rather than accurately observed as is the case with GPS. This situation has changed dramatically in recent decades due to the advances of geospatial technologies. Subsequently, large amounts of geospatial data at an individual level (rather than an aggregate level) have been collected for spatial analysis and spatial decision making related to events such as traffic jams, traffic accidents, and street crime. Network spatial analysis provides a powerful analytical means with which to gain insights into the data; for more details, one can refer to Okabe and Sugihara (2012). The network spatial analysis is not limited to street networks and can be applied to other networks such as river networks, water pipeline networks, and power networks. The network spatial analysis, accompanied by the SANET software tool (http://sanet.csis.u-tokyo.ac.jp/), is definitely a new addition to standard network analysis in GIS.

Although current network analysis in GIS is based on a graph in which street junctions and segments are represented by nodes and links, it is essentially geometric in the sense of locations of the junctions, distances of the segments, and/or directions between the segments or the junctions. This geometric representation of street networks, which is rooted and commonly seen in the geography literature (Hagget and Chorley 1969), is very helpful for dealing with such real-world problems as routing, closest facilities, and service areas – this is because the key issue for these problems is distance. However, the geometry-oriented network is of little help in addressing issues regarding the underlying structure of streets (note that, by streets, we mean entire named streets rather than street segments). For example, what is the average degree of street connectivity for a city? How does street connectivity differ from one city to another? How many intermediate streets must one pass in order to reach from street A to B? To address any of these issues, we need a topological representation or a graph in which entire named streets are represented by nodes, while street intersections or junctions are viewed as links. The graph, rooted in space syntax (Hillier and Hanson 1984) and related software tools (e.g., http://fromto.hig.se/~bjg/axwoman/), is purely topological, as it does not involve any geometric properties such as locations, distances, and directions (Jiang and Claramunt 2004). Many critics are skeptical of the topological representation, arguing that the geometric representation contains more metric information, and is therefore better than the topological one (e.g., Ratti 2004). This is a misperception, as having more information does not ensure that new insights will be obtained. On the contrary, having more information at the detailed level could prevent us from understanding the bigger picture, e.g., the scaling pattern of far more small things than large ones. We will return to this point below.

The topological representation is much more interesting and informative than the geometric one in terms of understanding the underlying structure. There is only a very limited range of connectivity for either junctions or segments, but a very wide range of connectivity for individual streets. In other words, there are only a few kinds of junctions or segments, but many kinds of streets. More importantly, the few kinds of segments are more or less similar and can be characterized by an average and a limited variance; the many kinds of streets, on the other hand, lack an average and a limited variance for characterizing their distribution of connectivity – this is known as scale-free, or scaling. To be more specific and for junctions in particular, the lowest number of connections is three, while the highest number could be around ten; the ratio of the highest to the lowest is about three or four. However, the picture is quite different for individual streets: the lowest and the highest connections are respectively one and dozens, so the ratio of the highest to the lowest is dozens. This large ratio indicates that there are far more less-connected streets than well-connected ones. In more general terms, there are far more small things than large ones, a universal scaling pattern observed in many natural and societal phenomena (Bak 1996). The geometric representation is of little help for uncovering the scaling pattern. In fact, the power of the topological representation lies exactly on the lack of geometric or metric information. In this regard, the London underground map adopted the same principle, being geometrically distorted and topologically retained, yet more informative in terms of station-to-station connections. As a reminder, a major purpose of building up models is to get a simplified counterpart (or model) of reality in order to obtain new insights into the reality. While there is no doubt that the reality itself contains far more information than the model, the reality can never be the model.

The scaling pattern, or the notion of far more small things than large ones, differs radically from that of more small things than large ones. The former indicates a nonlinear relationship – that is, a small cause large effect or large cause small effect – while the latter is a linear relationship – that is, small cause small effect, and large cause large effect. Street networks, or built environments in general, bear this scaling property, and are therefore nonlinear complex systems, which cannot be simply understood by Newtonian physics (Jiang 2014). In this connection, chaos theory and complexity science provide a series of tools such as fractal geometry, complex networks and agent-based modeling for better understanding the nonlinearity and complexity (Mitchell 2011). It is important to note that the scaling pattern of far more small things than large ones recurs multiple times, rather than just once. In other words, there are a few intermediate scales between the smallest and the largest, and all the scales form a scaling hierarchy. This scaling hierarchy can be derived by applying head/tail breaks – a new classification scheme for data with a heavy-tailed distribution (Jiang 2013). The resulting number of classes or hierarchical levels is called the ht-index (Jiang and Yin 2014), which can be used to characterize complexity of geographic features or fractals in general. The higher the ht-index, the more complex the fractal is.

We are not going to summarize the five papers, as they speak for themselves. We would like to take this opportunity to thank all the participants at the workshop for their active participation and discussions, the papers' authors for contributing their excellent works, and the many reviewers (who will be formally acknowledged separately by the journal) for providing the constructive review feedback in a timely fashion. We also would like to thank Dr. Xiaobai (Angela) Yao and Dr. Itzhak Benenson for their assistance in making the workshop successful and fun, Dr. Daniel Griffith, and Dr. Yongwan Chun, the journal editorial team, for their trust and support in editing the special issue.

**References**
Bak, P. (1996). *How Nature Works: The Science of Self-Organized Criticality*. New York: Springer-Verlag:.
Gil, J. (2014). "Analyzing the configuration of integrated multi-modal urban networks." *Geographical Analysis*, 46(4), 368-391.
Haggett, P., and R. Chorley. (1969). *Network Analysis in Geography*. London: Edward Arnold.
Hillier, B., and J. Hanson. (1984). *The Social Logic of Space*. Cambridge: Cambridge University Press.
Jiang, B. (2013). "Head/tail breaks: A new classification scheme for data with a heavy-tailed distribution." *The Professional Geographer* 65(3), 482–494.
Jiang, B. (2014). "Geospatial analysis requires a different way of thinking: The problem of spatial heterogeneity." *GeoJournal*, xx(x), xx-xx, DOI: 10.1007/s10708-014-9537-y.
Jiang, B., and C. Claramunt. (2004). "Topological analysis of urban street networks." *Environment and Planning B: Planning and Design* 31, 151–162.
Jiang, B., and J. Yin. (2014). "Ht-index for quantifying the fractal or scaling structure of geographic features." *Annals of the Association of American Geographers*, 104(3), 530–541.
Lerman, Y., Rofè, Y., and Omer, I. (2014). "Using space syntax to model pedestrian movement in urban transportation planning." *Geographical Analysis*, 46(4), 391-410.
Mitchell, M. (2011). *Complexity: A Guided Tour*. London: Oxford University Press.
Mohajeri, N., and Gudmundsson, A. (2014). "The evolution and complexity of urban street networks," *Geographical Analysis*, 46(4), 345-367.
Okabe, A., and K. Sugihara. (2012). *Spatial Analysis along Networks: Statistical and Computational Methods*. Chichester, UK: Wiley.
Ratti, C. (2004). "Space syntax: some inconsistencies." *Environment and Planning B: Planning and Design* 31, 487–499.
Shiode, S., and Shiode N. (2014). "Micro-scale prediction of near-future crime concentrations with street-level geo-surveillance." *Geographical Analysis*, 46(4), 435-455.
Wei, X., and Yao, X. (2014). "The random walk value for ranking spatial characteristics in road networks." *Geographical Analysis*, 46(4), 411-434.